\documentstyle[epsf]{article}

\title{
Nonequilibrium-current-induced corrections to
the one-particle-correlation function in a wire
}
\author{Yoshihiro Nishiyama \\
{\it Department of Physics, Faculty of Science,
Okayama University} \\
{\it Okayama 700-8530, Japan}
}
\date{(Received \hspace*{50mm})}

\begin{document}

\maketitle

E-mail: nisiyama@psun.phys.okayama-u.ac.jp

TEL: +81-86-251-7809

FAX: +81-86-251-7830

\section*{Abstract}
Electron gas in a wire connected to two terminals with
potential drop is studied with the Schwinger-Keldysh formalism.
Recent studies, where the current is enforced to flow
with a Lagrange-multiplier term, demonstrated
that the current
enhances the one-particle-correlation function.
We report 
that in our model, such enhancement is not guaranteed to occur,
but conditional both on the potential drop and the 
positions where we observe the correlation function.
That is, under a certain condition, 
spatially modulated pattern is formed in the wire
owing to the nonequilibrium current.

{\bf PACS codes(keywords):}
05.70.Ln (Nonequilibrium thermodynamics, irreversible processes),
72.      (Electronic transport in condensed matter),
85.30.Vw (Low-dimensional quantum devices (quantum dots, quantum wires, etc.))

Recently, Antal {\it et al.} have reported that
nonequilibrium current enhances the one-particle-correlation
function
contrary to our naive expectation that nonequilibrium transport
would destroy any orders \cite{Antal97,Antal98,Antal98b}.
They utilized a trick with which the nonequilibrium flow
is introduced in a frame of {\em equilibrium} statistical
mechanics.
That is, they introduced a Lagrange multiplier $\lambda$,
and 
investigated the (equilibrium) ground state
of the Hamiltonian given by
${\cal H} -\lambda J$, where
${\cal H}$ denotes a model Hamiltonian of a wire and
$J$ denotes 
the space
integral of the current operator.
Namely, the current flow is regarded as an environment
variable, which is to be given {\it ad hoc}.
As for an explicit example of ${\cal H}$, 
they took up
the transverse Ising model and the $XY$ model in one dimension,
and found that the above-mentioned behavior, namely,
the transport-induced correlation enhancement, occurs in these models.
Cardy extended their work for a nonintegrable case
(a lattice scaler field theory), and 
confirmed their observation \cite{Cardy99}.
This enhancement due to nonequilibrium
transport had been studied extensively
in the field of {\it classical} statistical mechanics
\cite{Garrido90,Cheng91,Schmittmann93,Praestgaard94,Bassler94,Bassler95}.
It should be noted, however, that the quantum version has an advantage
that the (nonequilibrium) dynamics is governed by the
Hamiltonian itself, whereas the classical dynamics depends upon
the kinetic rules which we implement.

In this letter, we investigate the spinless electron gas 
in a wire,
which is connected to two leads with different chemical potentials.
Hence, in our case, the current is driven to flow by the
potential drop as in actual experimental situations.
We stress that we have employed the Schwinger-Keldysh formalism
\cite{Schwinger61,Keldysh64},
that allows us to treat the transport phenomena driven far
out of equilibrium.
Thereby,
we report that the enhancement due to nonequilibrium 
flow
is not guaranteed to occur, but
depends significantly both on the potential drop
and the positions where we observe
the correlation.
As in the previous studies --- note that the transverse Ising model
and the $XY$ model are equivalent to {\em free}-%
spinless-fermion models ---
we supposed that the wire electrons are free (quadratic).
We believe that the many-body-correlation effect would not
change our essential conclusions.

Our model Hamiltonian is given by,
\begin{equation}
{\cal H}=
 \sum_k \left( \epsilon_k - \frac{eV}{2} \right) L^\dagger_k L_k
+\sum_k \left( \epsilon_k + \frac{eV}{2} \right) R^\dagger_k R_k
+\sum_i t (c^\dagger_i c_{i+1} + c^\dagger_{i+1} c_i)
+T'(c^\dagger_\alpha c_1+c^\dagger_1 c_\alpha)
+T'(c^\dagger_{\alpha'} c_L+c^\dagger_L c_{\alpha'})  .
\end{equation}
The operator $L^\dagger_k$ ($R^\dagger_k$) creates
a conduction electron with wave number $k$ in the left (right) lead.
$\epsilon_k$ denotes the dispersion relation of these leads, and
$\mu=eV/2$ ($-eV/2$) give the chemical potential of the left (right) lead.
Therefore, the parameter $eV$ gives the chemical-potential drop 
between the leads. 
The scheme how we describe nonequilibrium flow is explicated afterwards.
The operator $c_\alpha$ ($c_{\alpha'}$) is
the annihilation operator at the left (right) terminal;
namely, $c_\alpha=(1/\sqrt N)\sum_k L_k$
and $c_{\alpha'}=(1/\sqrt N)\sum_k R_k$ 
with the number of conduction
electron levels $N$.
The operator $c^\dagger_i$ creates an electron 
at site $i$ in the wire.
We assume that the wire is sufficiently long, and
the lead electrons are injected at the intermediate positions
of $i=1$ and $L$ with the coupling (transfer) amplitude $T'$.
We could implement the open-boundary condition for our wire,
for instance,
with 
use of the Fabrizio-Gogolin 
Tomonaga-Luttinger-liquid formalism under the open-boundary condition
\cite{Fabrizio95}; 
it would just cause inessential complications.

Properties of quantum wire, especially, those on the electron conductance,
have been studied extensively so far.
It seems, however, that these studies have been rather inclined to the vicinity
of equilibrium
that can be managed within
the Kubo formula, and little attention has been paid to the situation
driven far out of equilibrium.
The
Schwinger-Keldysh formalism \cite{Schwinger61,Keldysh64}
circumvents that restriction.
Their formalism is demonstrated for nonequilibrium
steady flow through a wire by Caroli {\it et al.} explicitly
\cite{Caroli71}.
We follow this description in order to investigate the nonequilibrium
flow.
In their description, the nonequilibrium steady flow
is realized through infinite-%
time evolution from such initial state
where both leads are disconnected from the wire ($T'=0$), and
are staying at each ground state with different chemical potentials.
With respect to such time-evolved nonequilibrium state
$| \ \  \rangle$,
the following Green functions are defined;
\begin{eqnarray}
G^{\rm r}_{ij}(t) &=& - {\rm i} \theta(t) 
      \langle \{c_i(t),c^\dagger_j\} \rangle \\
G^{\rm a}_{ij}(t) &=&   {\rm i} \theta(-t) 
      \langle \{c_i(t),c^\dagger_j\} \rangle \\
F_{ij}(t) &=& - {\rm i} 
      \langle  [c_i(t),c^\dagger_j] \rangle .
\end{eqnarray}
They satisfy the following respective Dyson equations,
\begin{eqnarray}
G^{\rm r}_{ij}(\omega)&=&g^{\rm r}_{ij}(\omega)
      +g^{\rm r}_{ik}(\omega)\Sigma_{kl}G^{\rm r}_{lj}(\omega)   \\
G^{\rm a}_{ij}(\omega)&=&g^{\rm a}_{ij}(\omega)
      +g^{\rm a}_{ik}(\omega)\Sigma_{kl}G^{\rm a}_{lj}(\omega)   \\
        F_{ij}(\omega)&=&        f_{ij}(\omega)
          +g^{\rm r}_{ik}(\omega)\Sigma_{kl}F_{lj}(\omega)
          +f_{ik}(\omega)\Sigma_{kl}G^{\rm a}_{lj}(\omega)   ,
\end{eqnarray}
The small-letter ones are those for the (unperturbed) initial state,
and
the self energy is given by 
$\Sigma_{kl}=T'(\delta_{k\alpha}\delta_{1l}+
                \delta_{k1}\delta_{\alpha l}+
                \delta_{k\alpha'}\delta_{Ll}+
                \delta_{kL}\delta_{\alpha'l} )$ \cite{Caroli71}.
It is notable that the Dyson equation of $F_{ij}$ is rather
involved, and as a matter of fact, nonequilibrium
characteristics are included in $F_{ij}$.

Quantity to be investigated below is the 
one-particle-correlation function, which is given by
$
C_{ij}= \langle c^\dagger_i c_j+c^\dagger_j c_i \rangle.
$
This correlation function is the central concern of the 
previous papers \cite{Antal97,Antal98},
where it is represented in terms of the spin language; 
$\langle S^x_i S^x_j \rangle$.
In one dimension (Tomonaga-Luttinger liquid),
(long-range asymptotic forms of)
all the correlation functions are interlocked 
each other.
Hence, 
one particular correlation function, for instance,
the one-particle correlation, is sufficient to gain
the essential physics.
(For example, when the one-particle correlation is suppressed,
the density-density correlation develops instead.)
The one-particle-correlation function 
is expressed in terms of the Green function $F_{ij}$,
\begin{eqnarray}
\label{o-p_correlation}
C_{ij} 
       &=&-\frac{\rm i}{2} \left( F_{ij}(t=0)+ F_{ji}(t=0) \right)   \\
\label{o-p_correlation2}
       &=&-\int \frac{ {\rm d}\omega }{2\pi}
            \frac{\rm i}{2}
             \left( F_{ij}(\omega)+F_{ji}(\omega) \right)   .
\end{eqnarray}
Therefore, we need to know $F_{ij}$.
By means of successive use of the above Dyson equations,
$F_{ij}$ can be expressed in terms of the unperturbed Green functions.
Through a straightforward linear algebra,
we arrive at the following expression;
\begin{equation}
\label{YN_formula}
F_{ij} = f_{ij}+g^{\rm r}_{i1}T'F_{\alpha j}+g^{\rm r}_{iL}T'F_{\alpha'j}
           +f_{i1}T'G^{\rm a}_{\alpha j}+f_{iL}T'G^{\rm a}_{\alpha'j},
\end{equation}
with
\begin{eqnarray}
F_{\alpha j} &=&
\left\{
       (1-g^{\rm r}_{\alpha'\alpha'}T'^2g^{\rm r}_{LL})
         (g^{\rm r}_{\alpha\alpha} T'(f_{1j}+f_{11}T'G^{\rm a}_{\alpha j}
           +f_{1L}T'G^{\rm a}_{\alpha' j})+f_{\alpha\alpha}T'G^{\rm a}_{1j})
\right.  \nonumber \\
& & \left.
    +g^{\rm r}_{\alpha\alpha}T'^2g^{\rm r}_{1L}
      (g^{\rm r}_{\alpha'\alpha'}T'(f_{Lj}+f_{L1}T'G^{\rm a}_{\alpha j}
          +f_{LL}G^{\rm a}_{\alpha' j})+f_{\alpha'\alpha'}T'G^{\rm a}_{Lj})
\right\} \\
& & /\left\{
       (1-g^{\rm r}_{\alpha\alpha}T'^2g^{\rm r}_{11})
       (1-g^{\rm r}_{\alpha'\alpha'}T'^2g^{\rm r}_{LL})
       -g^{\rm r}_{\alpha\alpha}T'^2g^{\rm r}_{1L}
        g^{\rm r}_{\alpha'\alpha'}T'^2g^{\rm r}_{L1}
     \right\}     \\
%
F_{\alpha'j} &=&
\left\{
       (1-g^{\rm r}_{\alpha\alpha}T'^2g^{\rm r}_{11})
         (g^{\rm r}_{\alpha'\alpha'} T'(f_{Lj}+f_{L1}T'G^{\rm a}_{\alpha j}
           +f_{LL}T'G^{\rm a}_{\alpha'j})+f_{\alpha'\alpha'}T'G^{\rm a}_{Lj})
\right.  \nonumber \\
& & \left.
    +g^{\rm r}_{\alpha'\alpha'}T'^2g^{\rm r}_{L1}
      (g^{\rm r}_{\alpha\alpha}T'(f_{1j}+f_{11}T'G^{\rm a}_{\alpha j}
          +f_{1L}G^{\rm a}_{\alpha'j})+f_{\alpha\alpha}T'G^{\rm a}_{1j})
\right\} \\
& & /\left\{
       (1-g^{\rm r}_{\alpha\alpha}T'^2g^{\rm r}_{11})
       (1-g^{\rm r}_{\alpha'\alpha'}T'^2g^{\rm r}_{LL})
       -g^{\rm r}_{\alpha\alpha}T'^2g^{\rm r}_{1L}
        g^{\rm r}_{\alpha'\alpha'}T'^2g^{\rm r}_{L1}
     \right\}     \\
%
G^{\rm a}_{\alpha j} &=& g^{\rm a}_{\alpha\alpha}T'G^{\rm a}_{1j}   \\
G^{\rm a}_{\alpha'j}&=& g^{\rm a}_{\alpha'\alpha'}T'G^{\rm a}_{Lj}   \\
G^{\rm a}_{1j}&=&
  \frac{
      g^{\rm a}_{1j}(1-g^{\rm a}_{LL}T'^2g^{\rm a}_{\alpha'\alpha'})
        +g^{\rm a}_{1L}T'^2g^{\rm a}_{\alpha'\alpha'}g^{\rm a}_{Lj}
       }
       {
      (1-g^{\rm a}_{11}T'^2g^{\rm a}_{\alpha\alpha})
         (1-g^{\rm a}_{LL}T'^2g^{\rm a}_{\alpha'\alpha'})
      -g^{\rm a}_{1L}T'^2g^{\rm a}_{\alpha'\alpha'}
       g^{\rm a}_{L1}T'^2g^{\rm a}_{\alpha\alpha}
       }                                \\
G^{\rm a}_{Lj}&=&
  \frac{
      g^{\rm a}_{Lj}(1-g^{\rm a}_{11}T'^2g^{\rm a}_{\alpha\alpha})
        +g^{\rm a}_{1j}T'^2g^{\rm a}_{\alpha\alpha}g^{\rm a}_{L1}
       }
       {
      (1-g^{\rm a}_{11}T'^2g^{\rm a}_{\alpha\alpha})
         (1-g^{\rm a}_{LL}T'^2g^{\rm a}_{\alpha'\alpha'})
      -g^{\rm a}_{1L}T'^2g^{\rm a}_{\alpha'\alpha'}
       g^{\rm a}_{L1}T'^2g^{\rm a}_{\alpha\alpha}
       }                               .
\end{eqnarray}

Therefore, the
remaining task is to express the unperturbed green functions
explicitly.
As is seen above, informations of the terminals are 
involved
within the local Green functions at $\alpha$ and $\alpha'$.
Here, we assume that the spectral property at the
terminals is of the Lorentz type; namely,
\begin{eqnarray}
g^{\rm r}_{\alpha\alpha ,\alpha'\alpha'} (\omega)
                               &=& 1/(\omega+{\rm i}W)  \\
g^{\rm a}_{\alpha\alpha ,\alpha'\alpha'} (\omega)
                               &=& 1/(\omega-{\rm i}W) \\
F_{\alpha\alpha} (\omega)
             &=& -2{\rm i}W{\rm sgn}(\omega-eV/2)/(\omega^2+W^2) \\
F_{\alpha'\alpha'}(\omega)
             &=& -2{\rm i}W{\rm sgn}(\omega+eV/2)/(\omega^2+W^2) ,
\end{eqnarray}
with the band width of the conduction-electron spectrum $W$.

As for the wire Green functions, we used the following
expressions;
\begin{eqnarray}
G^{\rm a}_{ij} (\omega)  &=& \frac{2\pi{\rm i}\rho}{v_{\rm F}}
    {\rm e}^{-{\rm i}\frac{\omega}{v_{\rm F}}|i-j|} 
      {\rm e}^{ - |\omega|/\omega_{\rm c}} \\
G^{\rm r}_{ij} (\omega)  &=& -\frac{2\pi{\rm i}\rho}{v_{\rm F}}
    {\rm e}^{{\rm i}\frac{\omega}{v_{\rm F}}|i-j|}  
      {\rm e}^{ - |\omega|/\omega_{\rm c}} \\
F_{ij}         (\omega)  &=&-\frac{4\pi{\rm i}\rho}{v_{\rm F}}
        \cos(\omega(i-j)/v_{\rm F}){\rm sgn}(\omega)  
      {\rm e}^{ - |\omega|/\omega_{\rm c}} .
\end{eqnarray}
with the density of states $\rho=1/(2\pi)$.
Let us mention some notions about these expressions:
We derive these expressions for a one-dimensional free-electron gas,
assuming that the one-particle dispersion 
is linear,
and the Fermi points are shifted
towards the origin; 
$\omega=\pm v_{\rm F}k$.
(We will assume that the units of time and length are identical
(isotropic);
$v_{\rm F}=1$.)
These prescriptions are common to those of the bosonization scheme,
and thus the validity is guaranteed especially as far as
the low-energy physics is concerned.
Moreover, it is notable that
we readily treat a many-body-correlated wire,
just substituting the above expressions with those of 
Luther and Peschel \cite{Luther74}.
The high-energy-cut-off factor $\omega_{\rm c}$
stands for the band width of
the wire electron, and is related to the lattice constant 
$a$;
$\omega_{\rm c}=v_{\rm F} k_{\rm c}=v_{\rm F}\pi/a$.
In the Tomonaga-Luttinger liquid theory, final expressions are
obtained through setting
$k_{\rm c} \to \infty$, 
because one is motivated to know the long-wave-%
length physics, where the lattice constant is renormalized to
zero.
Here, we retain the cut-off factor, 
because our wire length is supposed to be finite, and the
length is to be measured by the lattice constant.
We set the lattice constant as the unit of length; $a=1$.

The above formulae complete our scheme to 
calculate the one-particle-correlation function
(\ref{o-p_correlation}).
In order to do the numerical integration of eq. (\ref{o-p_correlation2}),
we adopted the Romberg algorithm \cite{NR};
we proceeded the Romberg iteration
until the output result converges up to the tenth digit.
All the data presented below are calculated for the condition of
the lead-electron-band width $W=2$, 
the coupling amplitude $T'=0.5$ and the
wire length $L=20$.

\begin{figure}[htbp]
\begin{center}\leavevmode
\epsfxsize=17cm
\epsfbox{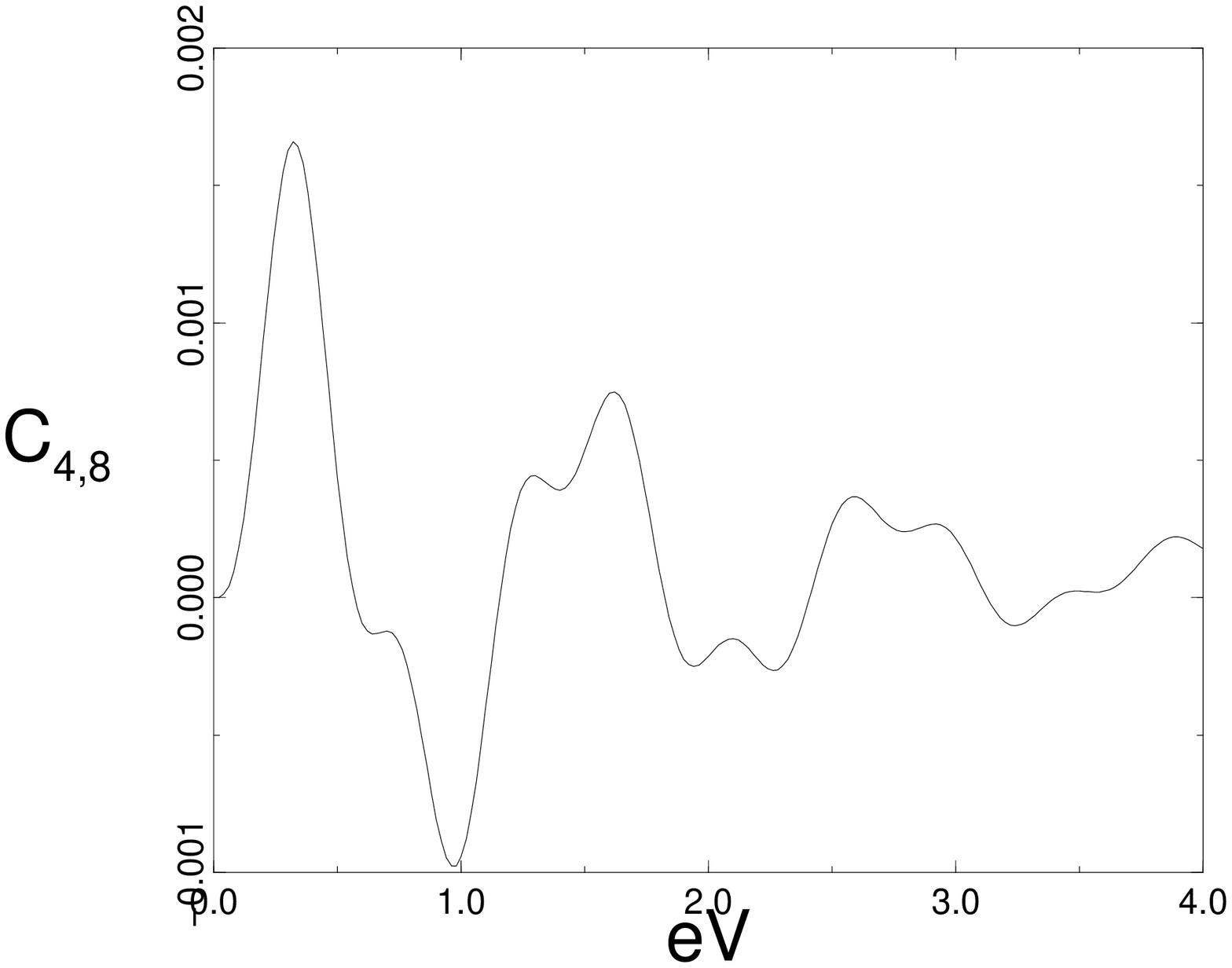}
\end{center}
\caption{
The one-particle-correlation function $C_{4,8}$
({\protect \ref{o-p_correlation}}) for various
chemical-potential drop $eV$.
}
\label{C_eV}
\end{figure}

Let us turn to present the numerical results.
In Fig. \ref{C_eV}, we plotted the correlation $C_{4,8}$
(\ref{o-p_correlation}) against the voltage drop $eV$.
According to the preceding reports, the correlation should be enhanced
by the nonequilibrium potential drop $eV$ \cite{Antal97,Antal98,Cardy99}.
We found that in fact, the correlation $C_{ij}$ is influenced
by $eV$.
However, the dependence is not monotonic, but
oscillates with respect to $eV$.
In other words, our result indicates that
it is not guaranteed whether the correlation
is stabilized or not, but conditional on $eV$.

In order to clarify the parameter range where this
non-equilibrium effect becomes significant,
we had swept various parameter ranges.
Thereby,
we observed that the oscillation period is not influenced
very much by $W$ and $T'$, but is proportional to $1/L$.
Hence, it is suggested that
the oscillation period is governed solely by the energy scale
$\sim \omega_{\rm c}/L$.
Note that the length of the wire does contribute to the
energy scale, and thus the existence of the terminals
is crucial in the present phenomenon.

\begin{figure}[htbp]
\begin{center}\leavevmode
\epsfxsize=17cm
\epsfbox{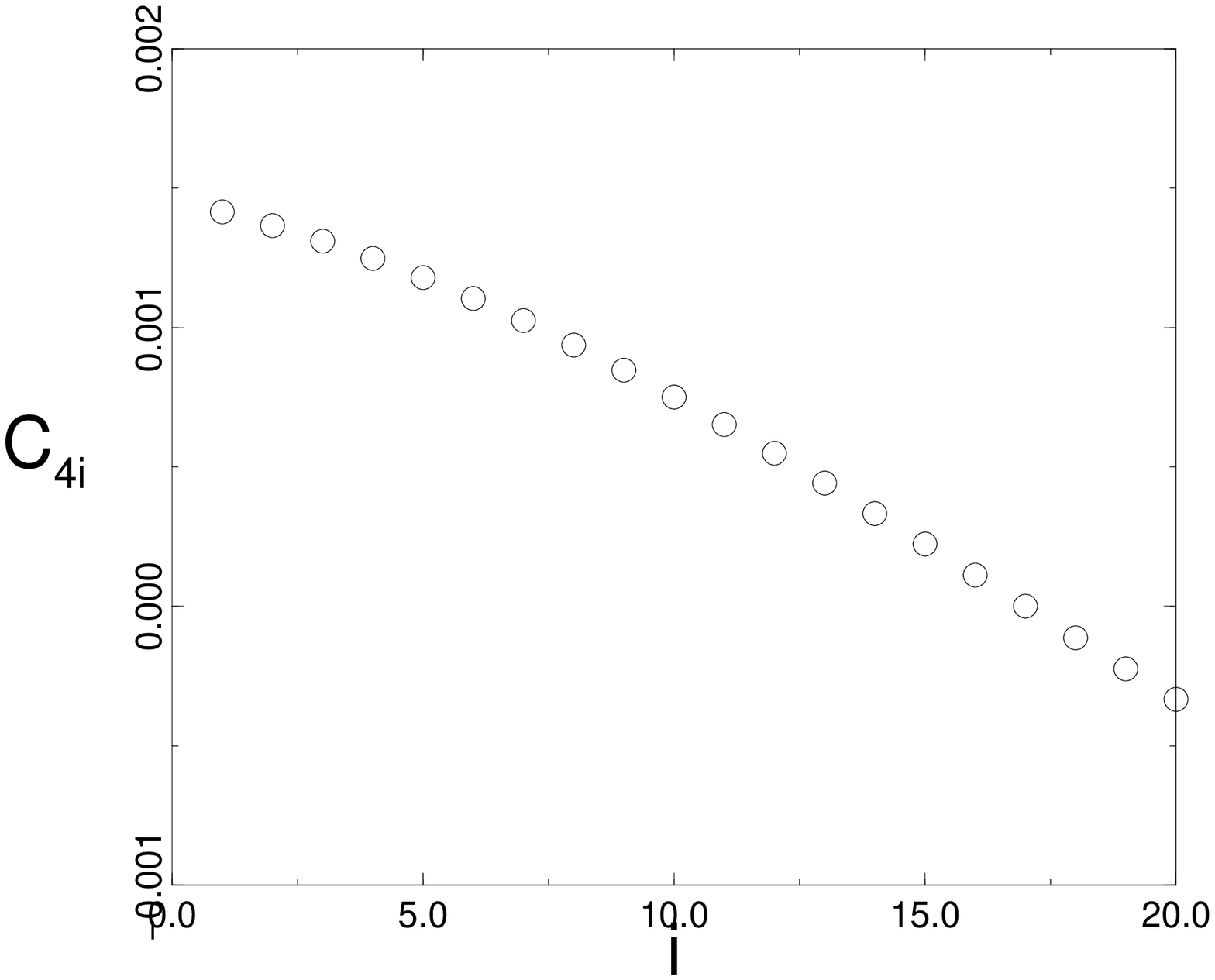}
\end{center}
\caption{
The one-particle-correlation function $C_{4,i}$
({\protect \ref{o-p_correlation}}) for various distances
$i$ over the wire for $eV=0.2$.
}
\label{C_i_a}
\end{figure}

\begin{figure}[htbp]
\begin{center}\leavevmode
\epsfxsize=17cm
\epsfbox{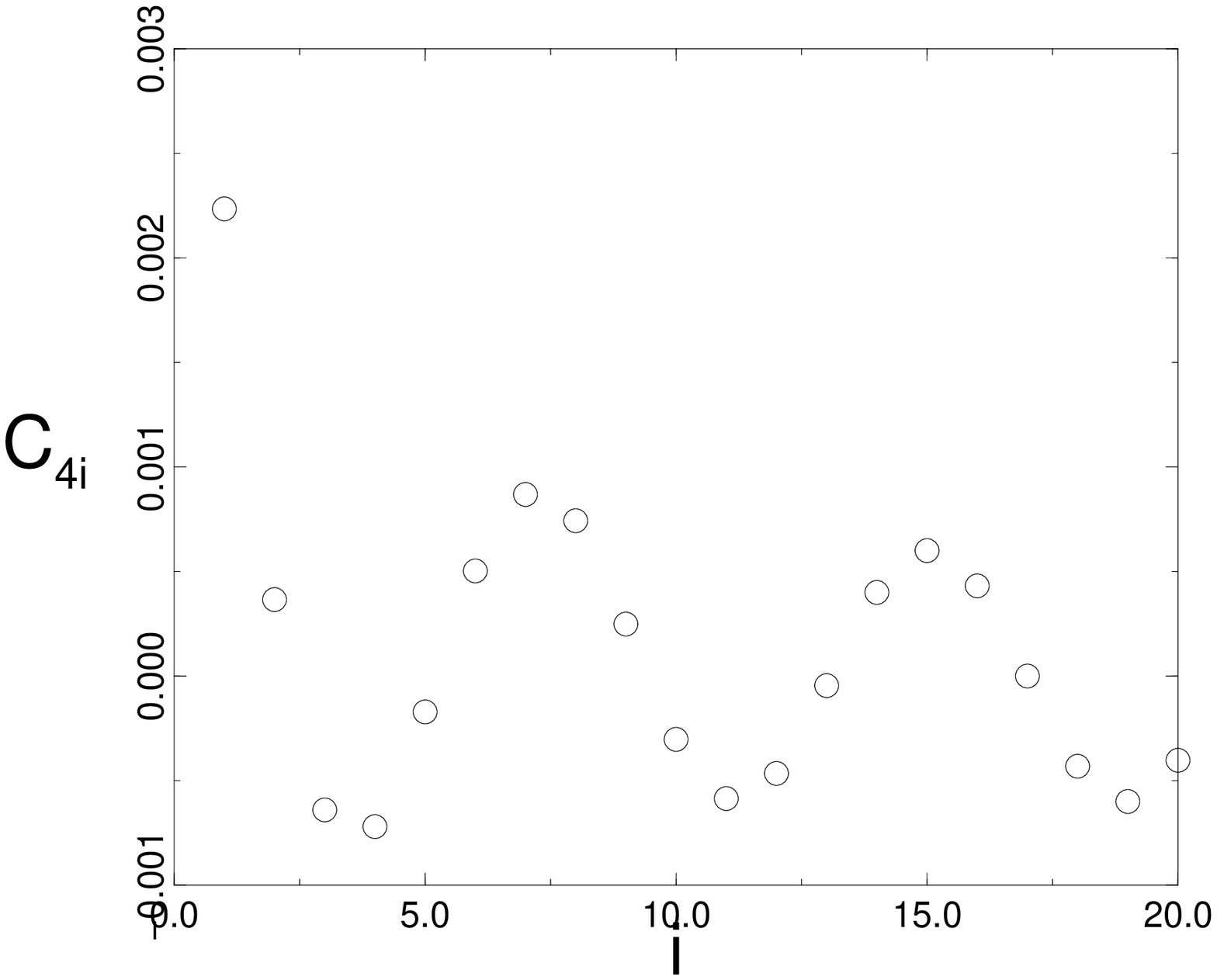}
\end{center}
\caption{
The one-particle-correlation function $C_{4,i}$
({\protect \ref{o-p_correlation}}) for various distances
$i$ over the wire for $eV=1.6$.
}
\label{C_i_b}
\end{figure}

Secondly, we report that the enhancement of $C_{ij}$
depends on the distance $i-j$ as well. 
In Figs. \ref{C_i_a} and \ref{C_i_b},
We plotted the correlation $C_{4,i}$ over various distances $i$
at $eV=0.2$ and $eV=1.6$, respectively.
At $eV=0.2$ (Fig. \ref{C_i_a}), 
we see that in almost whole distance range,
the correlation function decays monotonically, and
for long-range distance ($i-j \sim L$), to our surprise,
the sign of the correlation alternates.
That is, the correlation function suffers long-wave-length
modulation.
At $eV=1.6$ (Fig. \ref{C_i_b}), the alternation period becomes shortened.
We observed that for larger $eV$, in general, the correlation 
alternates more rapidly.
We stress that the modulation period
is not related to the (equilibrium) 
Fermi wave length like the Friedel oscillation,
because in our formalism the Fermi points are shifted towards $k=0$.
Hence, we conclude that the nonequilibrium current does
shift the Fermi points out of $k=0$ effectively.
The Fermi-point shift is found to take place actually in the
aforementioned models in refs. \cite{Antal97,Antal98}.
In Figs. \ref{C_i_a} and \ref{C_i_b}, we notice
that the correlation of $eV=1.6$ dominates that of $eV=0.2$.
One might think that the correlation is enhanced by the nonequilibrium
driving force $eV$.
Yet, we found that
this enhancement is rather accidental, and the correlation
amplitude becomes suppressed gradually by $eV$; remember the behavior
shown in Fig. \ref{C_eV}.

\begin{figure}[htbp]
\begin{center}\leavevmode
\epsfxsize=17cm
\epsfbox{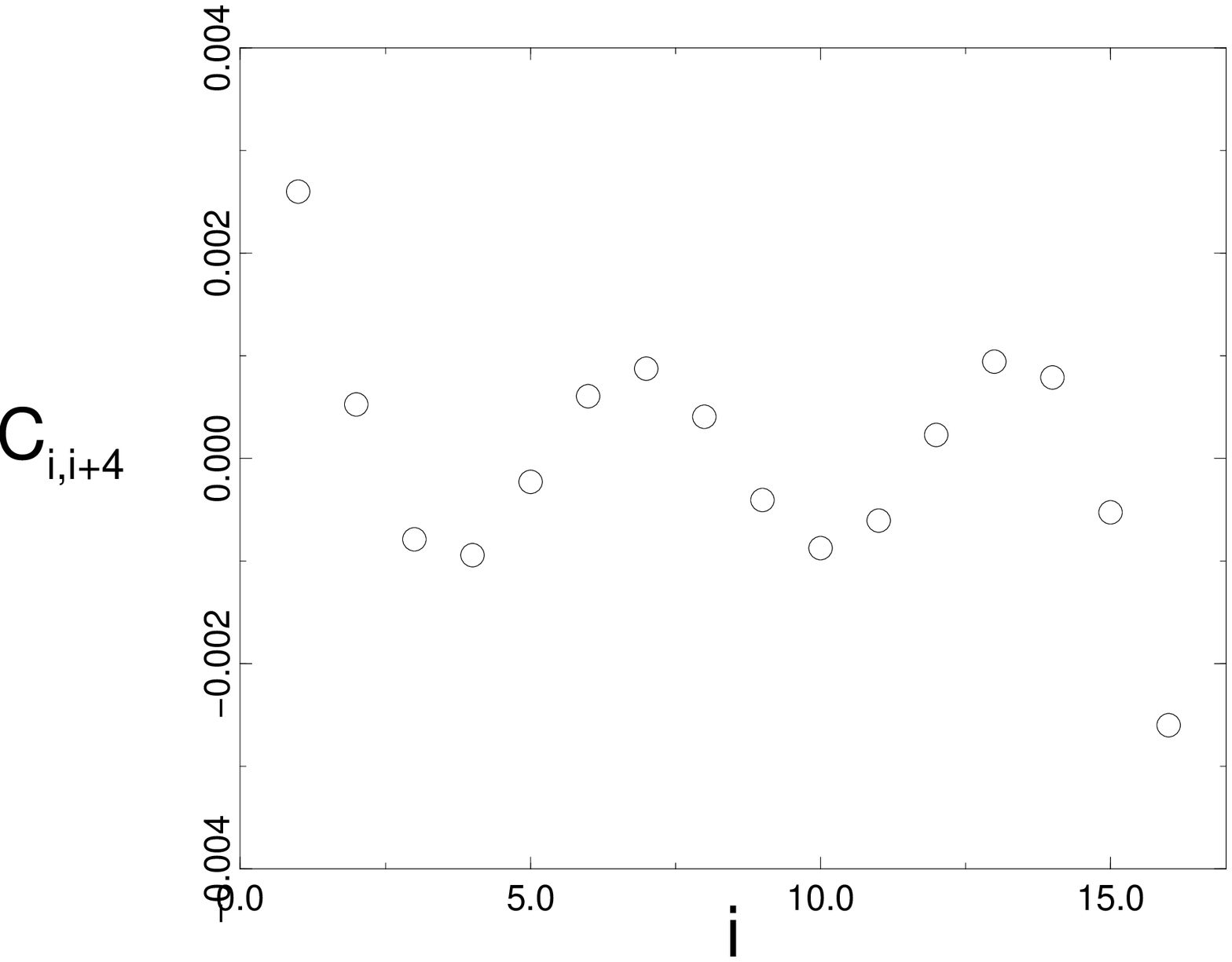}
\end{center}
\caption{
The one-particle-correlation function $C_{i,i+4}$
({\protect \ref{o-p_correlation}}) at various positions
$i$ in the wire for $eV=1$.
We see that the spatial distribution of the
correlation enhancement is anti-symmetric.
}
\label{C_ii}
\end{figure}

Next, we show the short-distance correlation $C_{i,i+4}$
for various positions at $eV=1$ in Fig. \ref{C_ii}.
We found that the correlation is spatially anti-symmetric
with respect to the center of the terminals.
Therefore, the correlation-function data
presented so far change the signs,
if we re-define $C_{ij}$ so as to become closer to the opposite
terminal instead; $C_{ij} \leftrightarrow C_{L-i+1,L-j+1}$.
This anti-symmetry may be plausible, if we notice the following
symmetry of
our model; $i \leftrightarrow L-i+1$ and $eV \leftrightarrow -eV$. 
Namely, owing to the anti-symmetry, the net enhancement of the correlation,
which is integrated over the wire, is always {\it vanishing}.
This feature is precisely due to the existence of terminals,
which are not taken into account explicitly in the previous studies.

\begin{figure}[htbp]
\begin{center}\leavevmode
\epsfxsize=17cm
\epsfbox{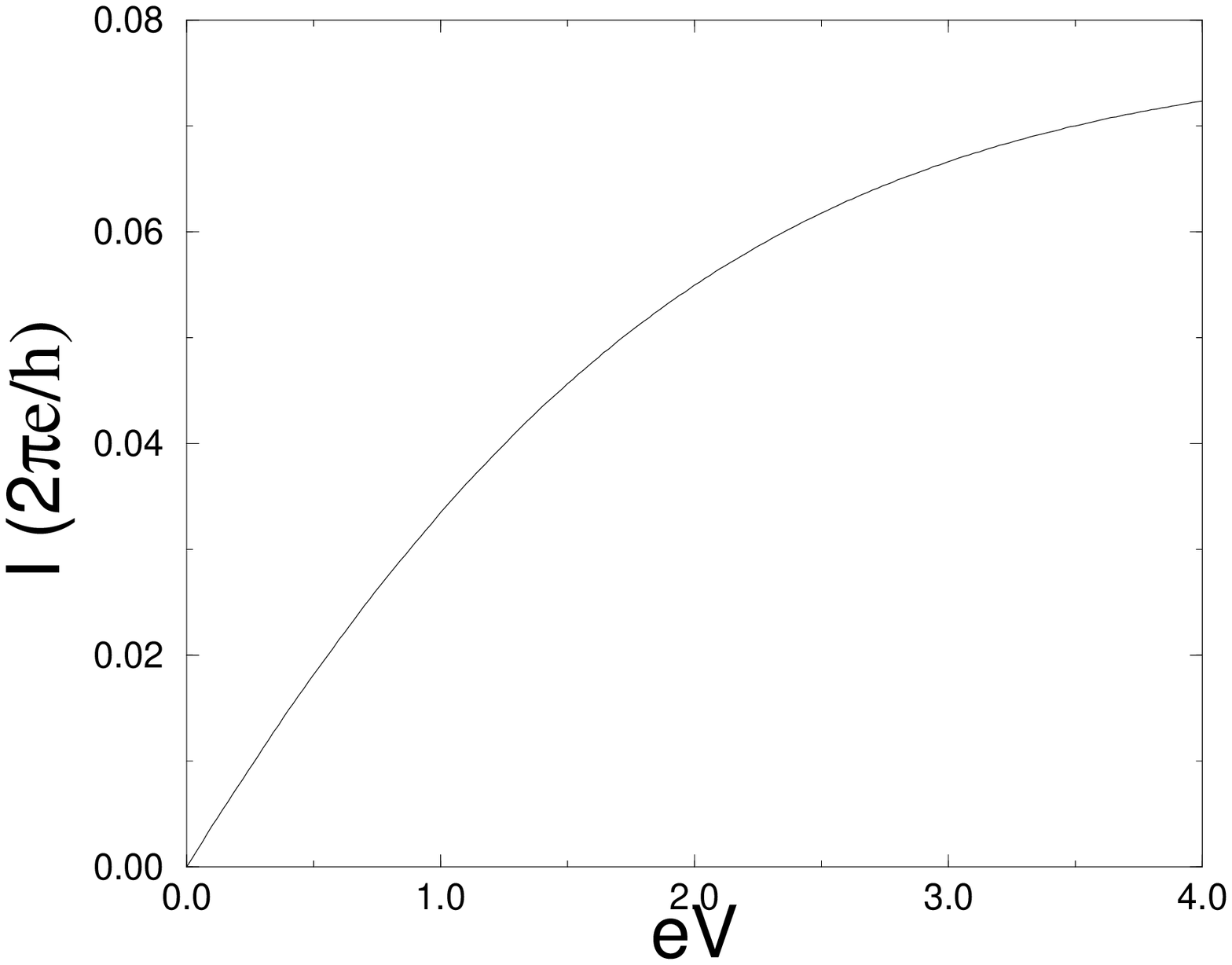}
\end{center}
\caption{
Electric current against the potential drop $eV$.
}
\label{denryu}
\end{figure}

Finally, In Fig. \ref{denryu},
we plotted the electronic current $I$ against $eV$.
The current is calculated by means of
the following formula derived by
Caroli {\it et al.} \cite{Caroli71};
\begin{equation}
I=\frac{eT'^4}{2\hbar}
\int \frac{{\rm d}\omega}{2\pi}
G^{\rm a}_{L1}G^{\rm r}_{1L}
\left\{
(g^{\rm a}_{\alpha\alpha}-g^{\rm r}_{\alpha\alpha})f_{\alpha'\alpha'}
+f_{\alpha\alpha}(g^{\rm r}_{\alpha'\alpha'}-g^{\rm a}_{\alpha'\alpha'})
\right\}  .
\end{equation}
We found that the dependence of the current on $eV$
is fairly monotonic; that does not exhibit any oscillatory
behaviors.
Hence, it is shown that our observations presented above
are not directly related to the current flow strength.

To summarize, we have calculated the one-particle-%
correlation function in a wire (\ref{o-p_correlation})
which is subjected to two biased terminals.
Being contrastive to the preceding report where
the nonequilibrium flow is introduced {\it ad hoc},
our results show that the enhancement of the correlation
function
due to the nonequilibrium flow is conditional 
both on the bias and 
the positions where we observe it.
In particular, at a certain condition, the nonequilibrium
flow gives rise to spatially modulated structures.
These features are related closely to the existence of terminals,
which are implicit in the previous Lagrange-multiplier formalism.
The situation that we had studied in this letter
is quite realistic, and it
may be realized in nano-technology devises.
However,
because we are concerned in the longstanding statistical-mechanical
problem whether
non-equilibrium transport enhances correlation or not,
we concentrated deeply on the one-particle correlation function, that
may not be measurable directly in experiments.
It remains for future study to calculate
quantities directly accessible in experiments.

\section*{Acknowledgments}
Hospitality at Institut f\"ur Theoretische Physik, Universit\"at
Hannover, is gratefully acknowledged.


\end{document}